\documentclass{ws-procs9x6-cpt22}
\usepackage{siunitx}
\usepackage[symbol]{footmisc}
\usepackage{wrapfig}

\begin{document}

\newcommand{\refeq}[1]{(\ref{#1})}
\def\etal {{\it et al.}}

\title{Understanding the gravitational and magnetic environment of a very long baseline atom interferometer}

\author{A.\ Lezeik,$^1$ D.\ Tell,$^{1}$, K. H. \ Zipfel$^{1}$, V. \ Gupta $^{1}$, E. Wodey$^{1}$\footnote[2]{Current affiliation: Alpine Quantum Technologies GmbH, Technikerstrasse 17/1, A-6020 Innsbruck, Austria}, E. M.\ Rasel$^1$, \mbox{C. \ Schubert$^{1,2}$,} D. \ Schlippert$^1$}

\address{$^1$Leibniz Universität Hannover, Institut für Quantenoptik, \\
Welfengarten 1, 30167 Hannover, Germany}

\address{$^2$German Aerospace Center (DLR), Institute for Satellite Geodesy and Inertial Sensing, Hannover, Germany}

\begin{abstract}
By utilizing the quadratic dependency of the interferometry phase on time, the Hannover Very Long Baseline Atom Interferometer facility (VLBAI) aims for sub \si{\nano\meter\per\square\second} gravity measurement sensitivity.
With its 10 \si{\meter} vertical baseline, VLBAI offers promising prospects in testing fundamental physics at the interface between quantum mechanics and general relativity. 
Here we discuss the challenges imposed on controlling VLBAI's magnetic and gravitational environment and report on their effect on the device's accuracy. 
Within the inner 8 \si{\meter} of the magnetic shield, residual magnetic field gradients expect to cause a bias acceleration of only $6\times10^{-14}$ \si{\meter\per\square\second} while we evaluate the bias shift due to the facility's non-linear gravity gradient to 2.6 \si{\nano\meter\per\square\second}.
The model allows the VLBAI facility to be a reference to other mobile devices for calibration purposes with an uncertainty below the 10 \si{\nano\meter\per\square\second} level.
\end{abstract}

\bodymatter

\section{Introduction}
An atom interferometer (AI) in a Mach-Zehnder configuration uses a sequence of three light pulses to measure different inertial quantities such as the local gravity, gravity gradients and rotations  \cite{kasevich1991PRL}. Excluding sensitivity to rotations by assuming a vertical motion of the atoms, the differential phase shift acquired between the two interferometer paths as a response to an acceleration $\vec{a}$ is proportional to the enclosed space-time area:
\begin{equation}
  \Delta\phi=\vec{k}_\mathrm{eff}\cdot \vec{a}  \,T^2\; ,
   \label{eq:phaseMZ}
\end{equation} 
where $\hbar \vec{k}_{\mathrm{eff}}$ is the effective momentum transferred during atom-light interaction and $T$ is the separation time between the light pulses.
The enclosed space-time area can be enhanced by  increasing the transferred momentum or by longer free fall times, achievable by performing experiments in microgravity~\cite{Becker2018} or by scaling up the size of the experiment, supporting a longer baseline~\cite{Johnson2011Thesis}. However, due to increased field variation, a 10 \si{\meter} long baseline as the one present at the Hannover Very Long Baseline Atom Interferometer (VLBAI) is far more prone to systematic bias phase shifts\cite{Sugarbaker2014Thesis,Wodey2021Thesis}.

The calculation of bias shifts caused by position-dependent potentials is
simplified by a perturbation theory approach for sufficiently small variations
of the potential. In such, a path integral over the full potential is evaluated
on trajectories that are solely determined by the unperturbed potential of
atoms in a constant gravitational potential\cite{Ufrecht2020PRA}.

In this article we focus on the magnetic field control \cite {Wodey2020RSI} and  the characterization of the gravitational field along the baseline \cite{Schilling2020JoG}  for accurate gravity measurements to be compared with low uncertainty to other gravimeters.

\section{Magnetic field control}
 
\begin{figure}[t]
\centering
\includegraphics[width=4.5in]{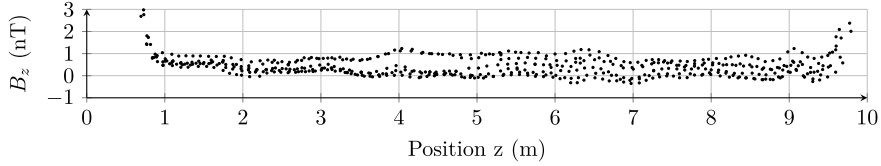}
\caption{Residual magnetic field along the baseline. The magnetic field gradient does not exceed 4 \si{\nano\tesla} over the central 8 \si{\meter} interferometry part of the baseline (ROI) with a longitudinal gradient of 2.5 \si{\nano\tesla\per\meter}. Adapted from Ref.\ \protect\refcite{Wodey2020RSI}.}
\label{fig:maggrad}
\end{figure}

One challenge for long baseline AI is to isolate the atomic cloud from non-inertial forces that affect the gravity measurement. 
In particular, spurious forces due to the coupling of the atoms to external magnetic fields generate bias accelerations on the atoms due to the Zeeman effect. AI experiments using alkali atoms therefore typically operate with $m_F=0$ states to avoid direct contribution due to the linear Zeeman effect. This reduces the Breit-Rabi formula, developed at second order in $B$ to $\Delta E=\frac{1}{2}\hbar\alpha B^2$,
where $\hbar$ is the reduced Planck's constant and $\alpha$ is the atomic species's clock transition Zeeman coefficient. Such a potential leads to a local bias acceleration along the z-direction on the atoms with mass $m$ and reads: 
\begin{equation}\label{eq:deltaa}
    \delta a= \epsilon\frac{\hbar\alpha B_0}{m}\frac{\partial b(z)}{\partial z} +\mathcal{O}(\epsilon^2)\; ,
\end{equation}
where we expanded $B(z)= B_0(1+\epsilon b(z))$, $\epsilon\ll1$. Therefore, a dual-layer octagonal magnetic shielding, made of Ni-Fe permalloy of high permeability, encloses the VLBAI facility's baseline to minimize the magnetic field magnitude to below 4 \si{\nano\tesla} and its longitudinal gradients to below 2.5 \si{\nano\tesla\per\meter} along the central 8 \si{\meter} of the baseline, the region of interest (ROI) (Fig.\ref{fig:maggrad}). In addition to shielding the baseline, a quantization axis needs to be realized to properly define a polarization axis of the transitions. 
This is achieved through winding a solenoid coil around the baseline vacuum tube to generate a field of $0.4$ \si{\micro\tesla\per\milli\ampere}. This field must be larger than any local variations to avoid depolarizing the atomic sample. Finally, we consider a constant bias field of $B_0=1.5$\si{\micro\tesla} and an AI within the ROI of the magnetic shield with $T=638$ \si{\milli\second} and zero initial velocity. Interpolating the measured magnetic field data and inserting it into eq. (\ref{eq:deltaa}),  we obtain from perturbation theory an acceleration bias of  $6\times10^{-14}$  \si{\meter\per\square\second}.\cite{Wodey2020RSI}

\section{Gravity profile along the baseline}\label{aba:sec1}
\begin{figure}[h]
\centering
\includegraphics[width=3.6in]{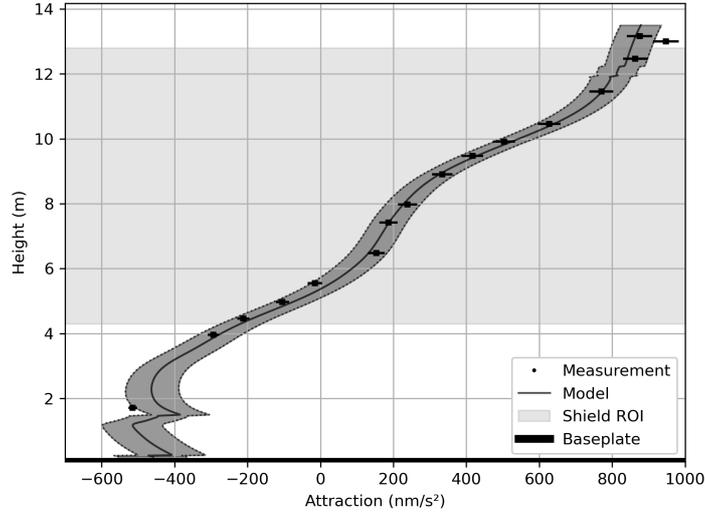}
\caption{Model and campaign results of the building's attraction ($\Delta g_{b,0} + \gamma_b z_\mathrm{eff} + \delta g_b(z)$) on the VLBAI baseline axis showing $95\%$ confidence (gray area). The light-gray area indicates the 8 \si{\meter} region of interest inside the magnetic shield. Adapted from Ref.\ \protect\refcite{Schilling2020JoG}.}
\label{fig:gravattraction}
\end{figure}

Operating a high resolution and absolute atom gravimeter requires taking a gravitationally inhomogeneous environment along the baseline into account. 
With a linear gravity gradient, the measured value of $a$, following from the phase shift via eq. \refeq{eq:phaseMZ}, corresponds to the value of local gravity $g$ only at a single position, the effective height $z_{\mathrm{eff}}$:
\begin{equation}
\begin{split}
    \Delta\phi_\mathrm{tot} &= k_\mathrm{eff}aT^2 = k_\mathrm{eff}(g_0-\gamma z_\mathrm{eff})T^2.
\end{split}
\label{eq:z_eff}
\end{equation}
$g_0=9.81$ \si{\meter\per\square\second} is the constant reference gravity value at $z=0$ measured at a baseplate located below the baseline\cite{Schilling2020JoG} and $\gamma$ is a linear gravity gradient.
In a Mach-Zehnder AI, the effective height is calculated as \cite{Peters2001Met}:
\begin{equation}
    z_\mathrm{eff} = z_0-\frac{7}{12}g_0T^2+\Bar{v}_0T,
\end{equation}
where $z_0=12.5$ \si{\meter} is the height at the start of the AI and $\Bar{v}_0=v_0+\frac{\hbar k_\mathrm{eff}}{2m}$ is the mean atomic velocity just after the AI opens ($v_0$ is the atomic velocity before the first light pulse and $m$ is the mass). In the VLBAI facility, the effective height of an AI within the ROI  ($T=638$ \si{\milli\second}, $v_0=0$) is \mbox{$z_\mathrm{eff}=11.6$ \si{\meter}}. For accurate gravity determination, the actual gravity variation along the baseline is considered.
It consists mostly of the linear gravity gradient, but contributions from local masses like ground density, building structure and heavy laboratory equipment also need to be taken into account.
Therefore, a full numerical model of the gravitational environment of the VLBAI facility was created \cite{schillingthesis} and validated by two measurement campaigns\cite{Schilling2020JoG}.
Decomposing every contribution into a constant and linear part, and gathering all non-linear residuals in a single perturbation term $\delta g(z)$ leads to the following description: 
\begin{equation}
 g(z)=(g_0+\Delta g_{\mathrm{g},0}+\Delta g_{\mathrm{b},0})-(\gamma_0+\gamma_{\mathrm{g}}+\gamma_{\mathrm{b}})z+\delta g(z)
 \label{eq:gtotal}
\end{equation}
with Earth's linear gravity gradient $\gamma_0=3.1$ \si{\micro\meter\per\square\second\per\meter}, and the constant and linear parts of the ground and building's attraction respectively $\Delta g_{\mathrm{g},0}=-7$ \si{\nano\meter\per\square\second}, $\gamma_{\mathrm{g}}=-219$ \si{\nano\meter\per\square\second\per\meter}, $\Delta g_{\mathrm{b},0}=-727$ \si{\nano\meter\per\square\second} and $\gamma_{\mathrm{b}}=-126$ \si{\nano\meter\per\square\second\per\meter}.
The latter is plotted against the VLBAI facility's height including the measured campaign values in Fig.(\ref{fig:gravattraction}). Finally, subtracting the linear gravity model from the full numerical model, we obtain $\delta g(z)$ being sufficiently small to perform the perturbation theory approach \cite{Ufrecht2020PRA}. Specifically, with $T=638$ \si{\milli\second}, $\delta g=+2.6$  \si{\nano\meter\per\square\second}  which needs to be added to the gravity value calculated at $z_\mathrm{eff}$ as shown in eq. \refeq{eq:z_eff} to obtain:
\begin{equation}
 g(z_\mathrm{eff})=\frac{\Delta\phi_\mathrm{tot}}{k_\mathrm{eff}T^2}+\delta g
 \label{eq:phitot}
\end{equation}

\section{Conclusion and outlook}
We have shown the control of the magnetic gradients and modeled the gravitational gradients along the baseline. On one hand, magnetic field gradients are shielded below 2.5 \si{\nano\tesla\per\meter} and contribute to the bias acceleration at the sub \si{\nano\meter\per\square\second}. On the other hand, gravity gradients need to be modeled to account for inhomogeneities. We have shown that the non-linearities $\delta g$ in the model lead to a bias acceleration of 2.6 \si{\nano\meter\per\square\second}. With the achieved degree of control, performance of an absolute atom gravimeter would only be limited by the uncertainty of the gravity model\cite{Schilling2020JoG}. By transferring the gravity values measured by VLBAI at $z_\mathrm{eff}$ to neighboring positions, the Hannover VLBAI facility offers the perspective of providing a reference and calibration station for mobile devices. Furthermore, we envisage applications for testing fundamental physics, e.g. quantum clock interferometry \cite{pumpoufrecht} and in searching for beyond-Riemann gravity~\cite{Kostelecky21PRD}.

\section*{Acknowledgments}
This work is funded by the Deutsche Forschungsgemeinschaft (DFG, German Research Foundation): Project-ID 274200144 - SFB 1227 DQ-mat (projects B07 and B09),  Project-ID 434617780 - SFB 1464 TerraQ (project A02), and Germany’s Excellence Strategy - EXC-2123 QuantumFrontiers - Project-ID 390837967 - and from “Niedersächsisches Vorab” through the “Quantum- and Nano-Metrology (QUANOMET)” initiative within the Project QT3. D.S. acknowledges support by the Federal Ministry of Education and Research (BMBF) through the funding program Photonics Research Germany under contract number 13N14875.

\end{document}